\documentclass[12pt,a4paper]{article}
\usepackage[utf8x]{inputenc}
\usepackage{graphicx}
\usepackage{authblk}
\usepackage{a4wide}

\title{Measurement of the Spectra of Single Bubble Sonoluminescence in water }
\author{J. Anto\v{s}}
\affil{Institute of Experimental Physics, SAS, Ko\v{s}ice, Slovakia}
\begin{document}

\maketitle

\begin{abstract}
  A preliminary results of measurement of the spectra of Single Bubble Sonoluminescence (SBSL) in water are presented. Analysis concentrates on
  similarity and differences of spectra from black-body radiation like shape. 
\end{abstract}
\section{Introduction}
Phenomenon of Single Bubble Sonoluminescence (SBSL) is known for over  quarter  of the century. Still some basic questions are not answered yet. 
 There is a consensus (at least in principle) on bubble hydrodynamics, expansion and implosion of gas bubble trapped in (standing) 
 ultrasonic field. As a consequence high temperature is expected inside a bubble at final stages of implosion. There remain differences
 about absolute value of a temperature and it's dependence on several parameters. Differences are on level of several orders of magnitude.
 There is a consensus that SBSL represents most efficient known energy focusing mechanism in physics \cite{nature}. Energy focusing by this 
 mechanism is (at least) 11 orders of magnitudes. Differences remain e.g. concerning question if one is able by this mechanism achieve temperatures 
 needed for nuclear fusion. There are published results claiming that this target was already achieved \cite{taler} however these results have been
  met with heated criticism.\\
 Analysis of spectra from SBSL is considered as a most effective approach to understand processes and conditions when SBSL is created. Already 
 the first measurements of spectra from SBSL produced surprising results. 
 Spectrum of SBSL resembled black-body radiation at 
 high temperature. Lack of line structure in the spectrum was suspected to be because temperature at SBSL is much higher
 than in case of e.g. Multi Bubble Sonoluminescence (MBSL) where OH lines have been identified \cite{spec}. Later same OH lines have been
 found also in ``dim'' SBSL \cite{specOH}. Spectroscopic study of SBSL in sulfuric acid brought important evidence
 about lines of excited states of Ar and the first evidence of existence of plasma in case of SBSL \cite{specAr}. Several other experiments
 detected Ar and Na lines \cite{specNaAr} or $Tb^{+3}$ lines \cite{specTb3,halide}, PO lines \cite{specPh}. Observation of spectral lines in SBSL
 spectra adds important piece of information about SBSL to a full picture.  
  Very important results about time evolution of the spectra from SBSL were presented in \cite{timeres}. Single flashes of SBSL were divided
  into 1 nano second
 bins time sequences and in each bin spectrum was measured. Shape of spectra in each bin resembles  black-body radiation with corresponding temperature rising from 
 few thousand in first couple of bins up to $\approx$ 100000 K in the last bin.
 This was specific measurement where working liquid was sulfuric acid and noble gas in which SBSL was created was Kr. But in a case this feature is general one there is
 a chance 
  to see deviation from single black-body radiation shape even in an integral spectra in SBSL with different working liquids and working gasses. \\
  There have been lot of measurements of SBSL when working liquid was a water in the past \cite{brenn}. In this paper we analyze results of a series of
  measurement of the spectra
  of SBSL in UV-VIS range 250 - 900 nm. We observe spectra which resemble  black-body radiation shape for a specific temperatures but also 
  spectra which are clearly impossible to describe by a single black-body radiation function.
  
  \section{Apparatus}
\begin{figure}[h]
%\emphasize 12 cm 
\begin{center}
\includegraphics[width=3in]{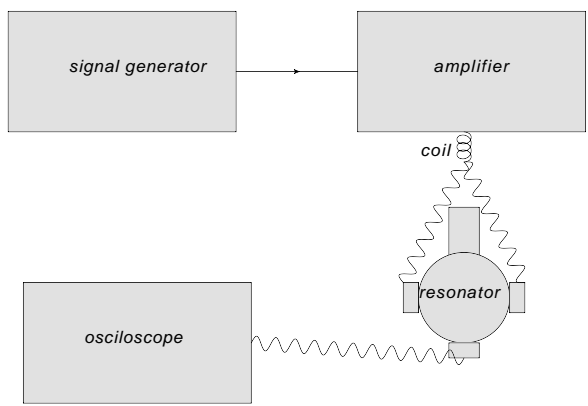} \\

\caption{schematic view of apparatus for SBSL
\label{SLsch}
} 
\end{center}

\end{figure}

  Apparatus consists of sort of standard equipment to produce SBSL (resonator, amplifier, signal generator - see Fig. \ref{SLsch}) and spectrometer QE65000 \cite{ooqe65} to
  measure spectra.
  Below we briefly  emphasize some details  we consider
   important for present analysis.

  \subsection{Resonator}
  Beaver flask component of resonator belongs to standard laboratory glass. We tested couple flasks made by different producers. In this paper results are based
  on  quartz beaver 100 ml flask resonator. Considered resonator is equipped with two piezoelectric transducers (type C5400 from Channel Industries, Inc.) glued there facing each other on equatorial side of beaver flask.
  Their purpose is to create standing 
  ultrasonic wave which traps gas (air) bubble when flask is filled with proper liquid. At the bottom of beaver flask is glued small piezoelectric 
  element which serves as a microphone. In our case a working liquid was a water. \\
  At a proper conditions (frequency and amplitude set in a signal generator) bubble oscillates, 
  expands and then implodes
  and  produces short burst of light per period. 
  \subsection{Interface of optical system and spectrometer}
\begin{figure}[h]
%\emphasize 12 cm 
\begin{center}
\includegraphics[width=3in]{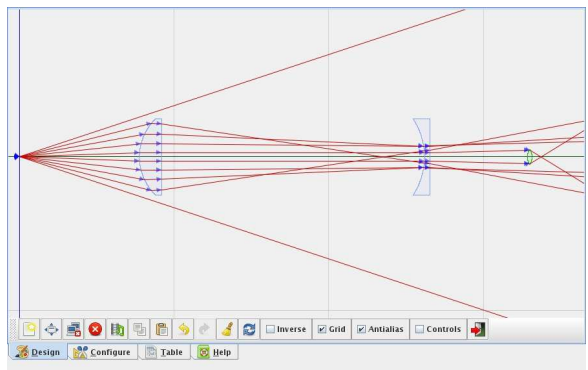} \\

\caption{Optical configuration for measurement of spectra of SBSL
\label{lensestd}
} 
\end{center}
\end{figure}

   Schema of optical configuration is displayed in Fig. \ref{lensestd}. Optical system consists of 3 lenses made from fussed silica with focal length
   $f_{L}$ = 38, -50 and 10 mm respectively.
   The last one is coupled to optical fiber through connector sma905 and other end of a fiber is through the same type of connector connected to a
   spectrometer. System was tuned in such a way that ``shining bubble'' was on optical axes of the above optical system. Optical system is placed
   as close as possible to resonator.
   
\subsection{Procedure for creation of SBSL}
\begin{figure}[h]
%\emphasize 12 cm 
\includegraphics[width=10cm]{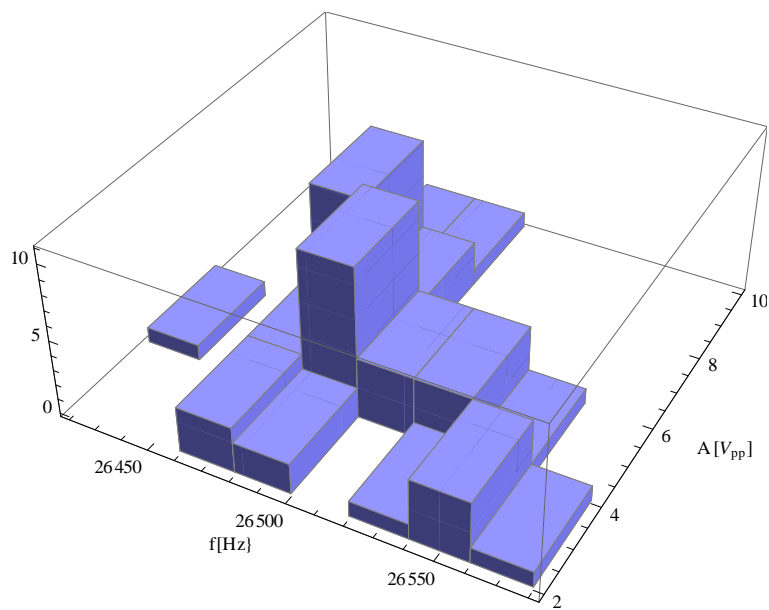} 
%\space* {}
\caption{A 3d histogram of starting frequencies x  amplitudes of signal generator when SBSL was achieved} 
\label{freqamp}
\end{figure}

 Stable SBSL, appropriate for measurement of spectrum (with our current apparatus), is produced  in degassed water only.
 Water was degassed by reducing atmospheric pressure in a vessel with water to $\approx$ 4 kPa and at the same time stirring it by magnetic stirrer.
  This procedure takes about 15-30 minutes. Level of degassing is checked by a measurement of level of dissolved oxygen in degassed water. Sample used
  for measurement of spectra of SBSL had at beginning of measurement dissolved oxygen content between 1.4 - 2.6 mg/l. Because resonator is opened
  to air level of degassing of water changes during a measurement.\\
  At proper frequency and amplitude
  of signal generator, air bubble which can be created in different ways (e.g. gently stirring surface of water in resonator) is captured
  at center of a resonator. By further tuning of amplitude and/or frequency bubble starts to shine and SBSL is created. A 3d plot
  of frequency (Hz)  x amplitude ($V_{pp}$) - parameters of the signal generator - conditions at which SBSL bubbles have been created is displayed in Fig. \ref{freqamp}.
 
  \subsection{Calibration of the spectrometer}
   Spectrometer is calibrated by procedure described in detail in \cite{calibSL}. Briefly, in the region 400 - 900 nm spectrometer is calibrated 
   by using calibration lamp \cite{ools1}, below 400 nm (up to 250 nm) calibration is done by extrapolating well defined black-body 
   radiation fit
   in region 400 - 900 nm  of (selected) SBSL measured sample down to 250 nm. One should note that this way is calibrated shape of distribution.
   There was no effort done to re-calibrate scale for transition from simple optical fiber to optical system as in Fig. \ref{lensestd}.
   \subsection{Measurements of the spectra}
\begin{figure}[h]
%\emphasize 12 cm 
\begin{center}$
\begin{array}{cc}
\includegraphics[width=2.5in]{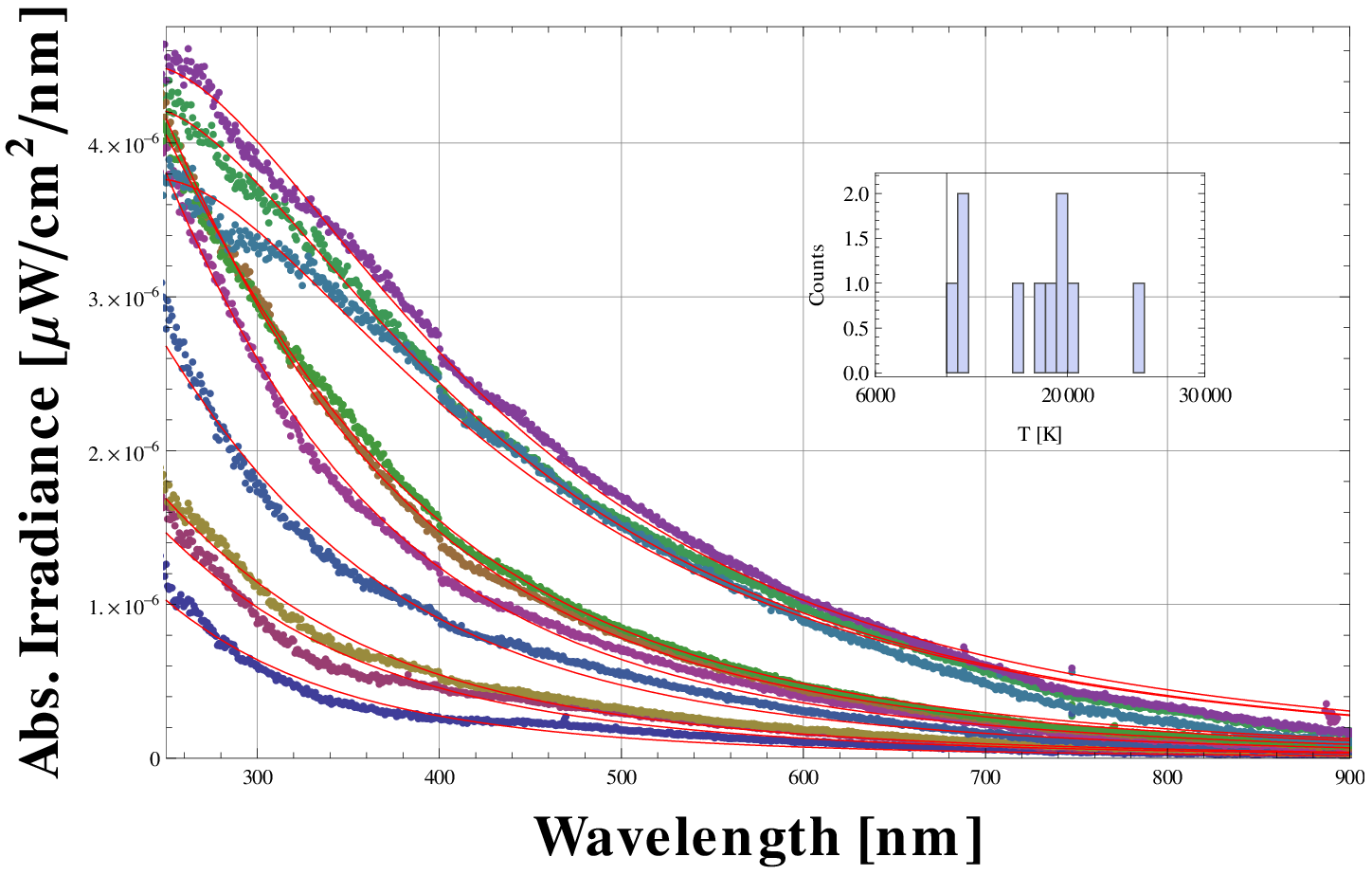} &
\includegraphics[width=2.5in]{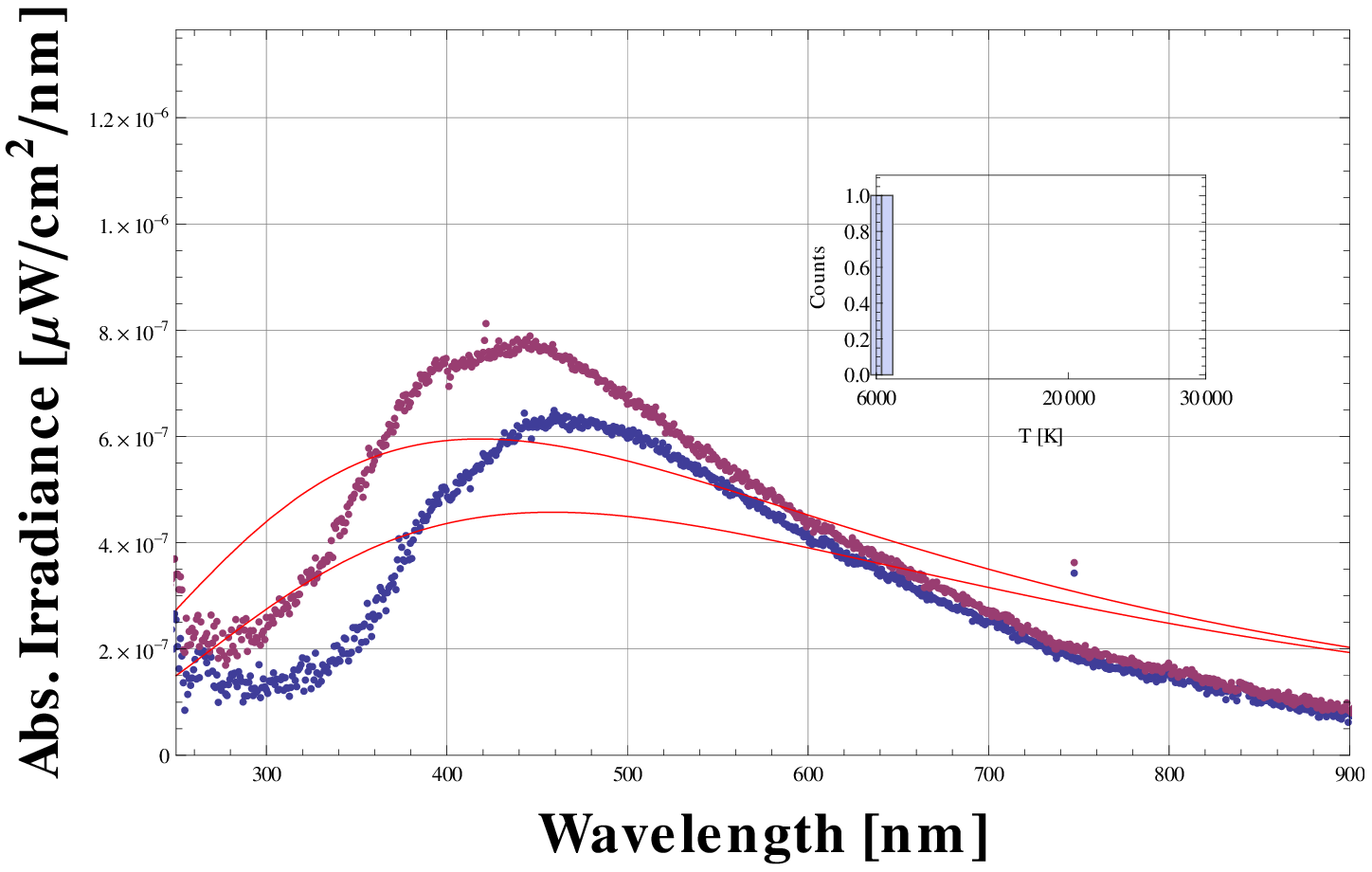} \\
\includegraphics[width=2.5in]{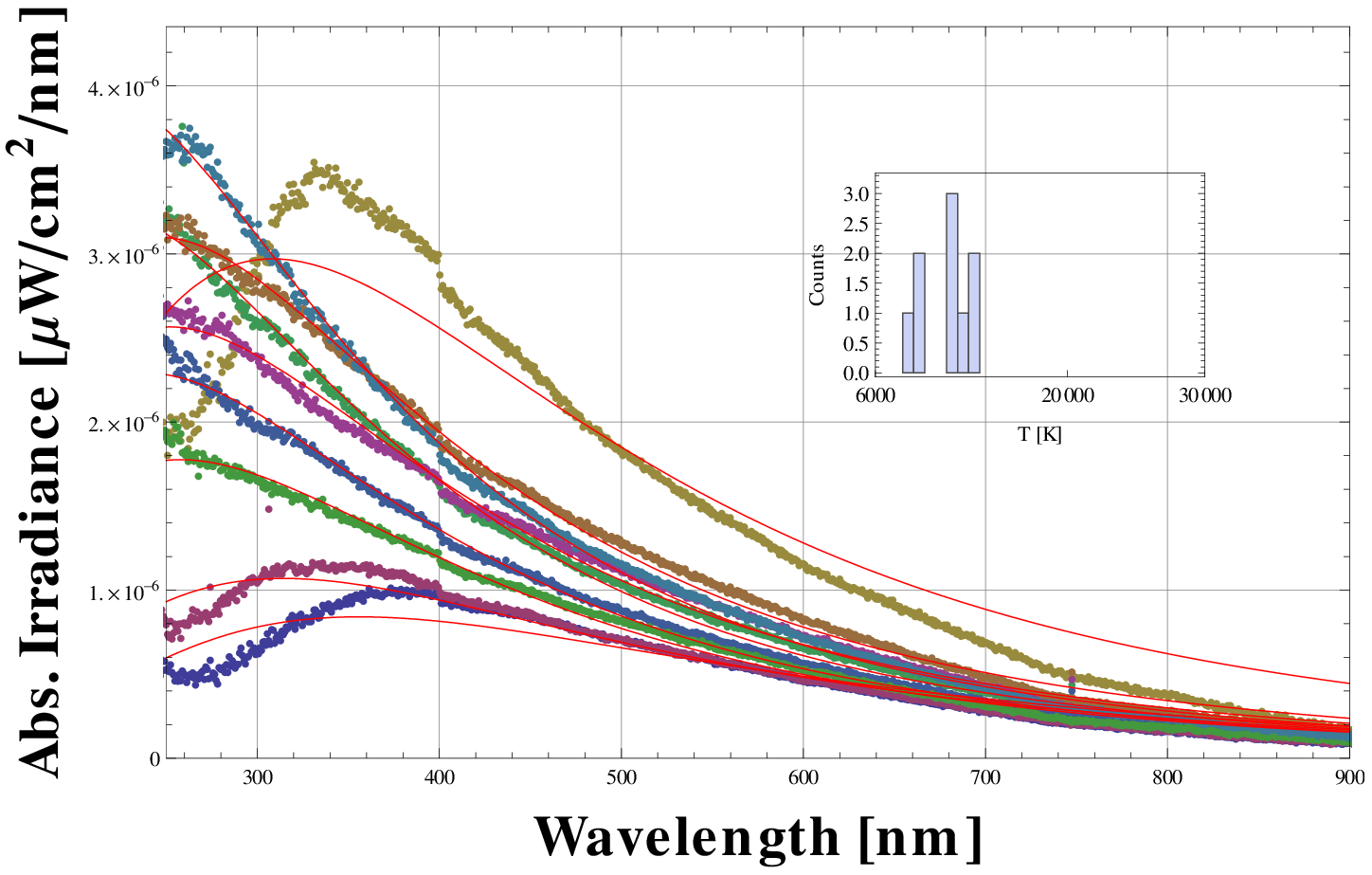} &
\includegraphics[width=2.5in]{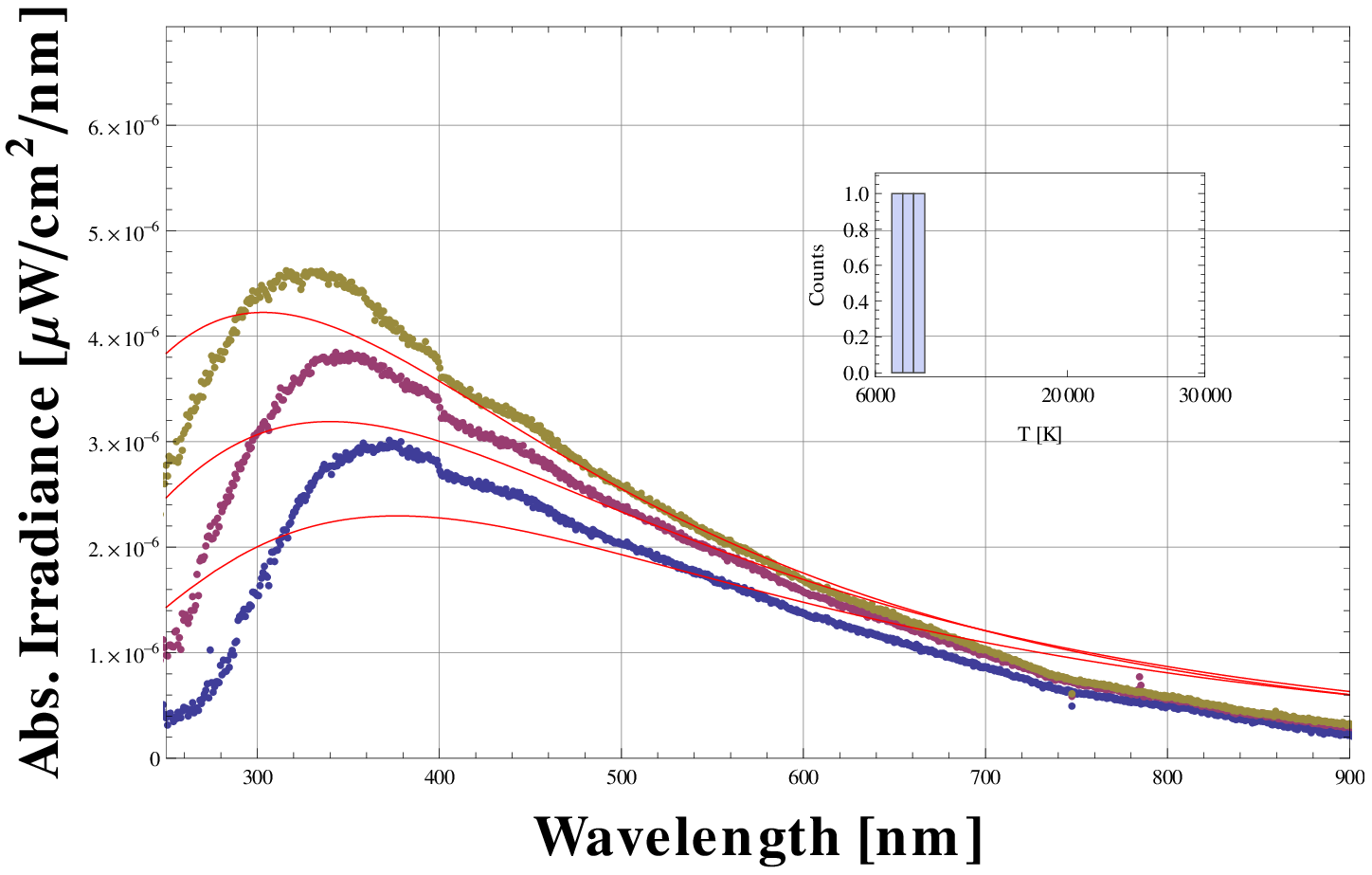} \\
\includegraphics[width=2.5in]{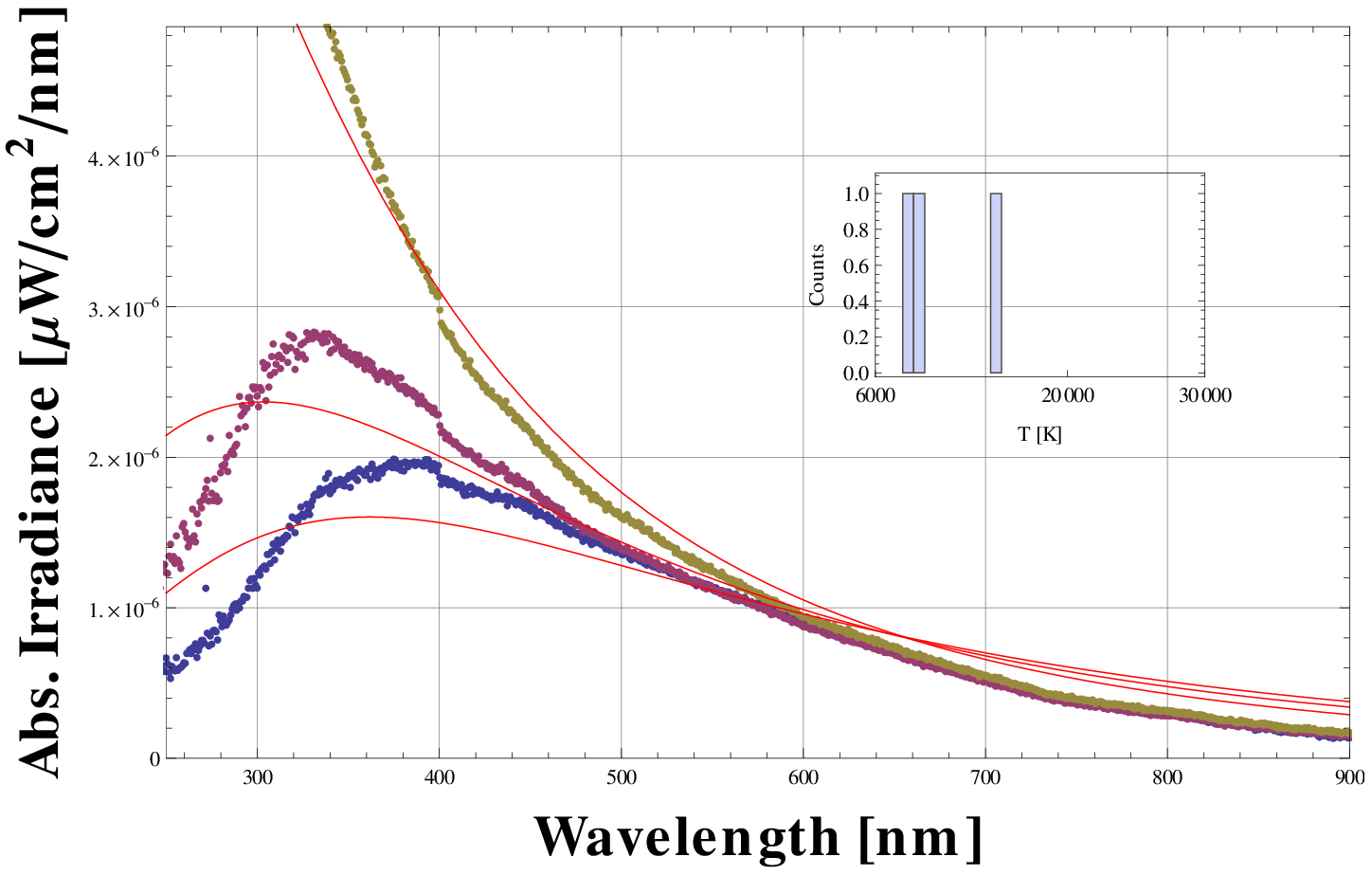} & 
\\
\end{array}$
\end{center}
\caption{ One can see that some spectra fit perfectly black-body radiation shape (red curve) in other cases comparison is quite rough.
Inset in each plot is histogram of fitted temperature of black-body radiation. Mean of temperature for above plots from top to bottom and from 
left to right is 17631, 6623, 11082, 8590, 10741 K.  } 
\label{UVabs1}
\end{figure}

\begin{figure}[h]
%\emphasize 12 cm 
\begin{center}$
\begin{array}{cc}
\includegraphics[width=2.5in]{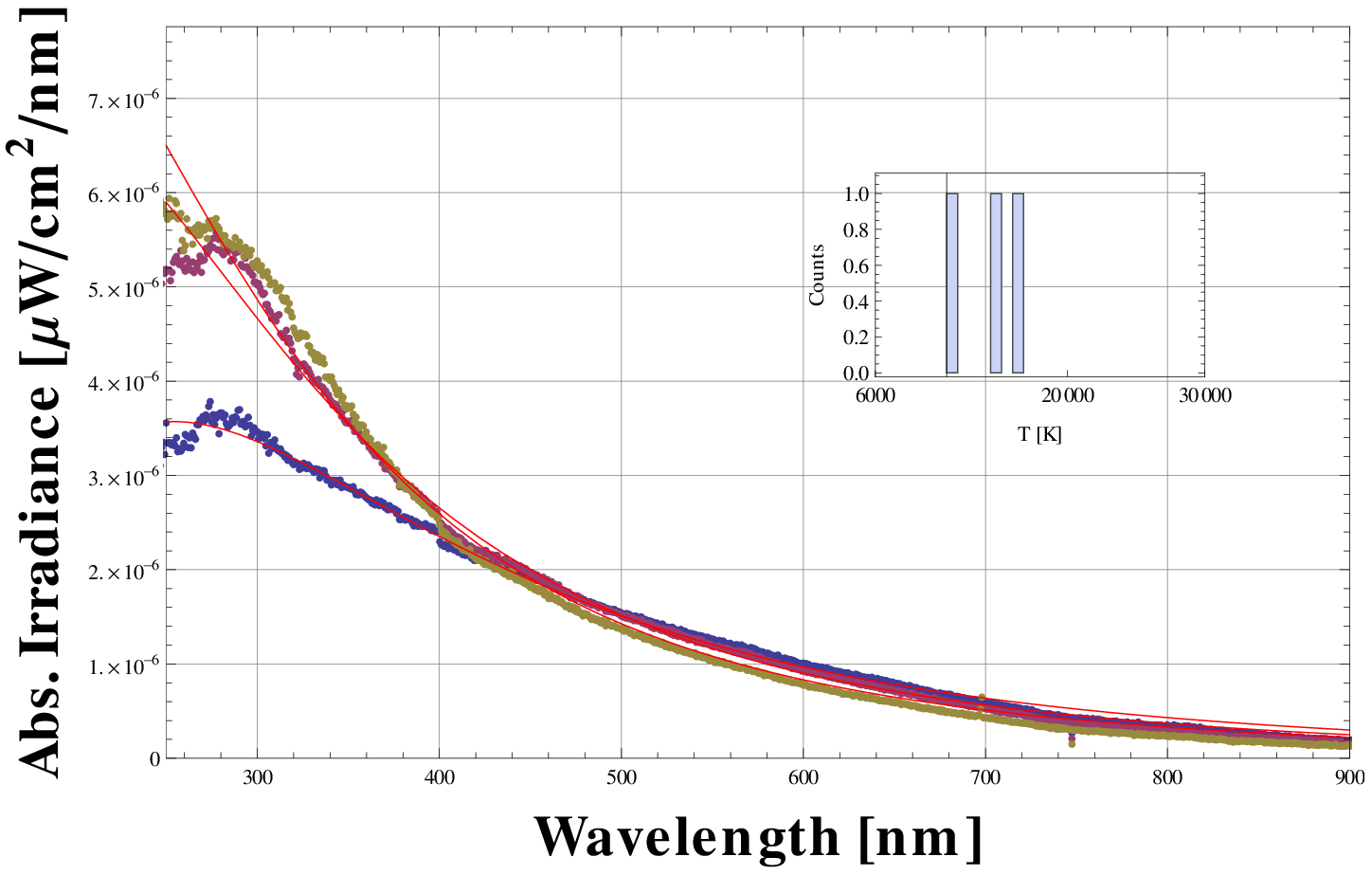} &
\includegraphics[width=2.5in]{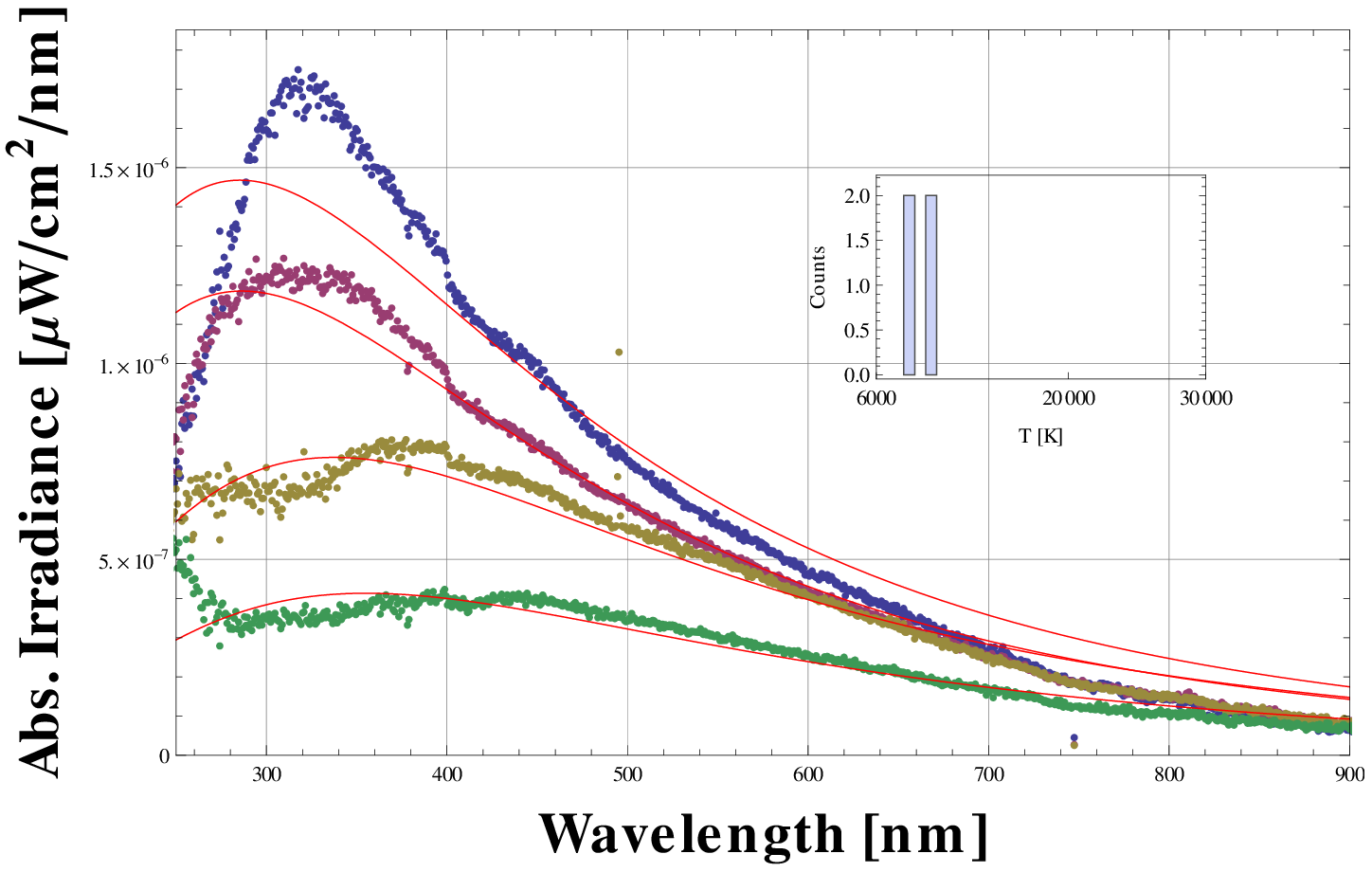} \\
\includegraphics[width=2.5in]{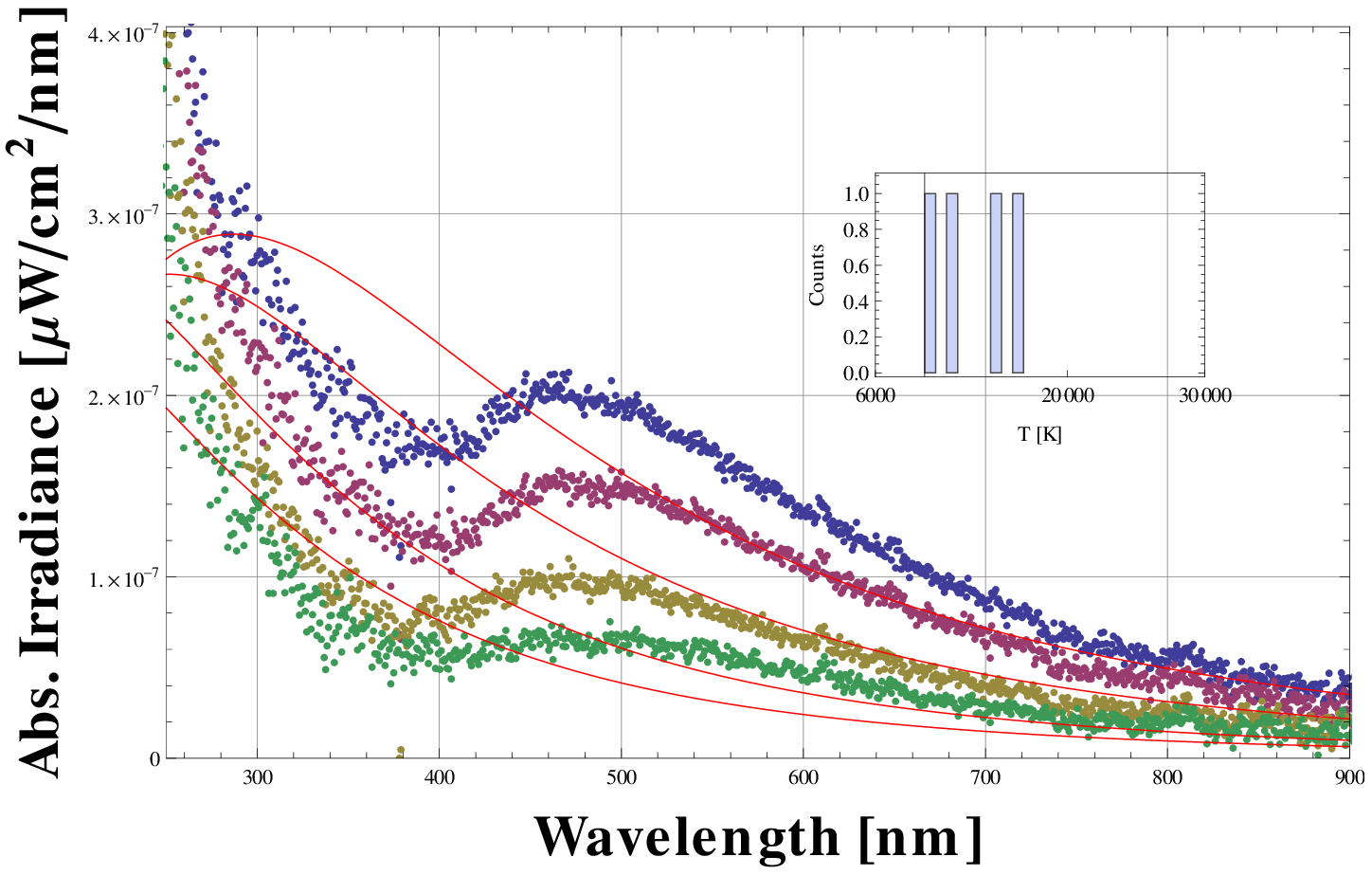} &
\includegraphics[width=2.5in]{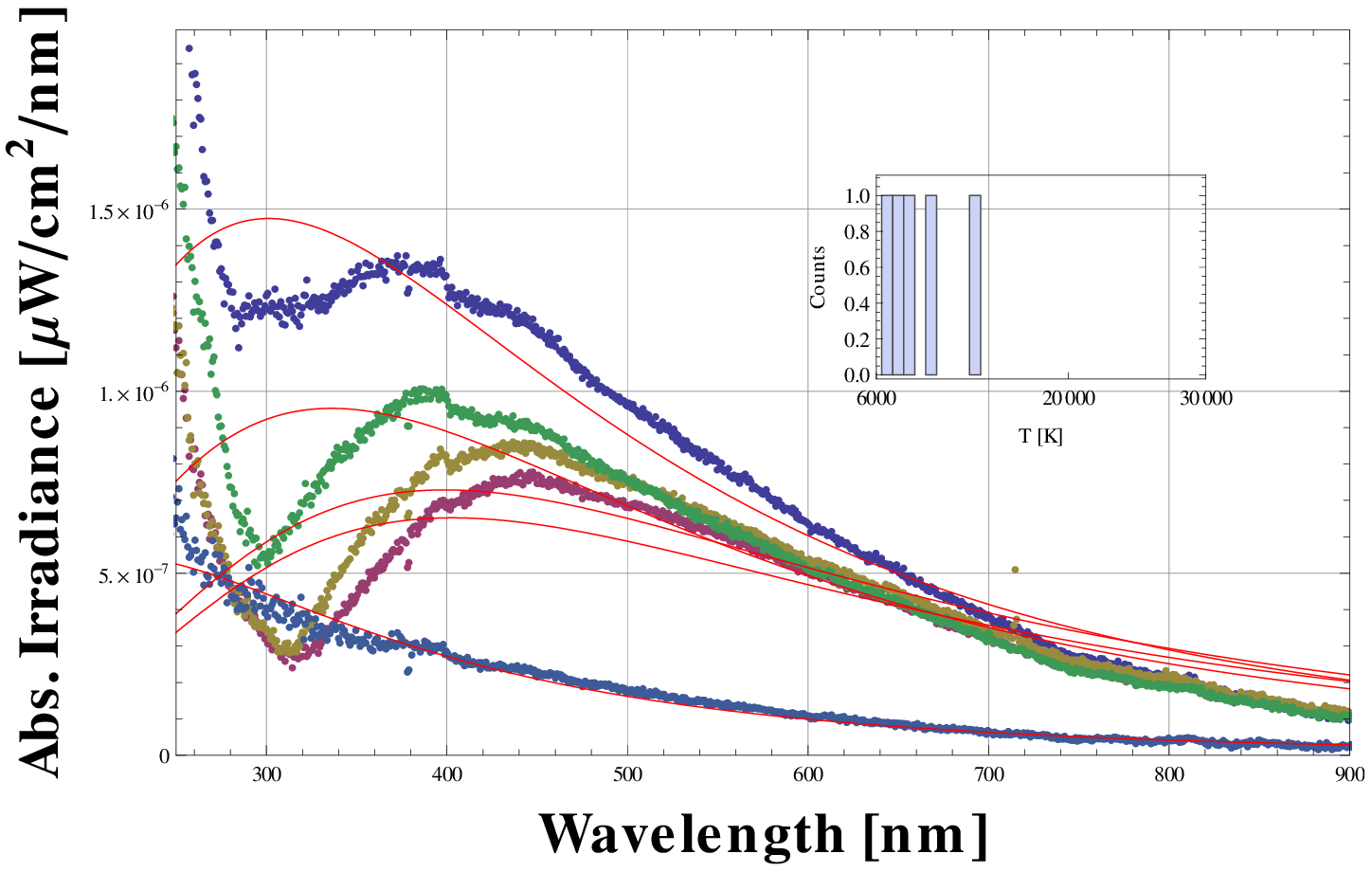} \\
\includegraphics[width=2.5in]{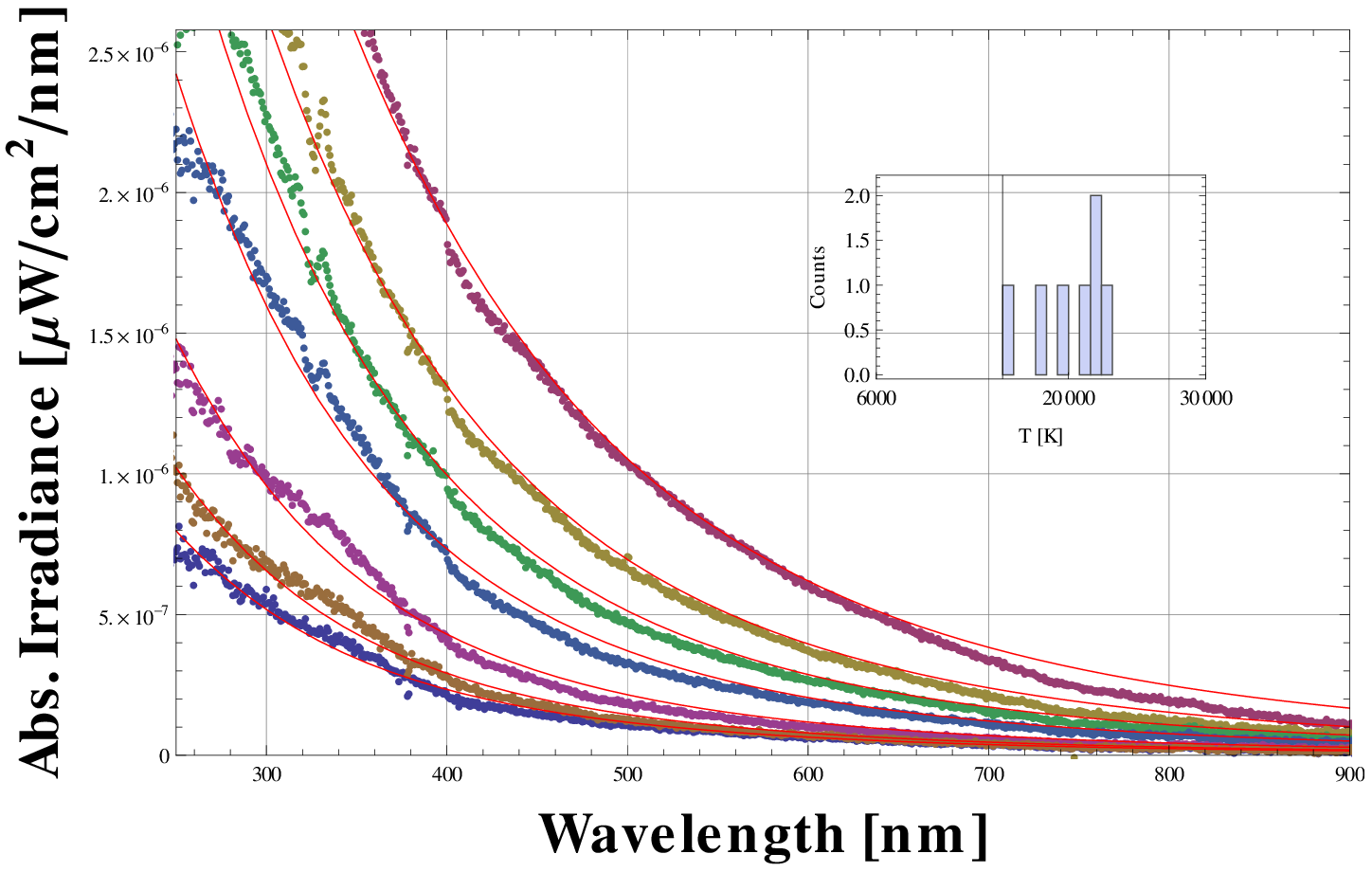} &
\\
\end{array}$
\end{center}
\caption{Same comments apply as for Fig. \ref{UVabs1}. Specially, in case of middle left plot - it looks like spectrum is composed of two sources of different temperature. 
 Mean of fitted temperature for above plots from top to bottom and from 
left to right is 13980, 9256, 13148, 9179, 20159 K.  } 
\label{UVabs2}
\end{figure}

\begin{figure}[h]

\begin{center}$
\begin{array}{cc}
\includegraphics[width=2.5in]{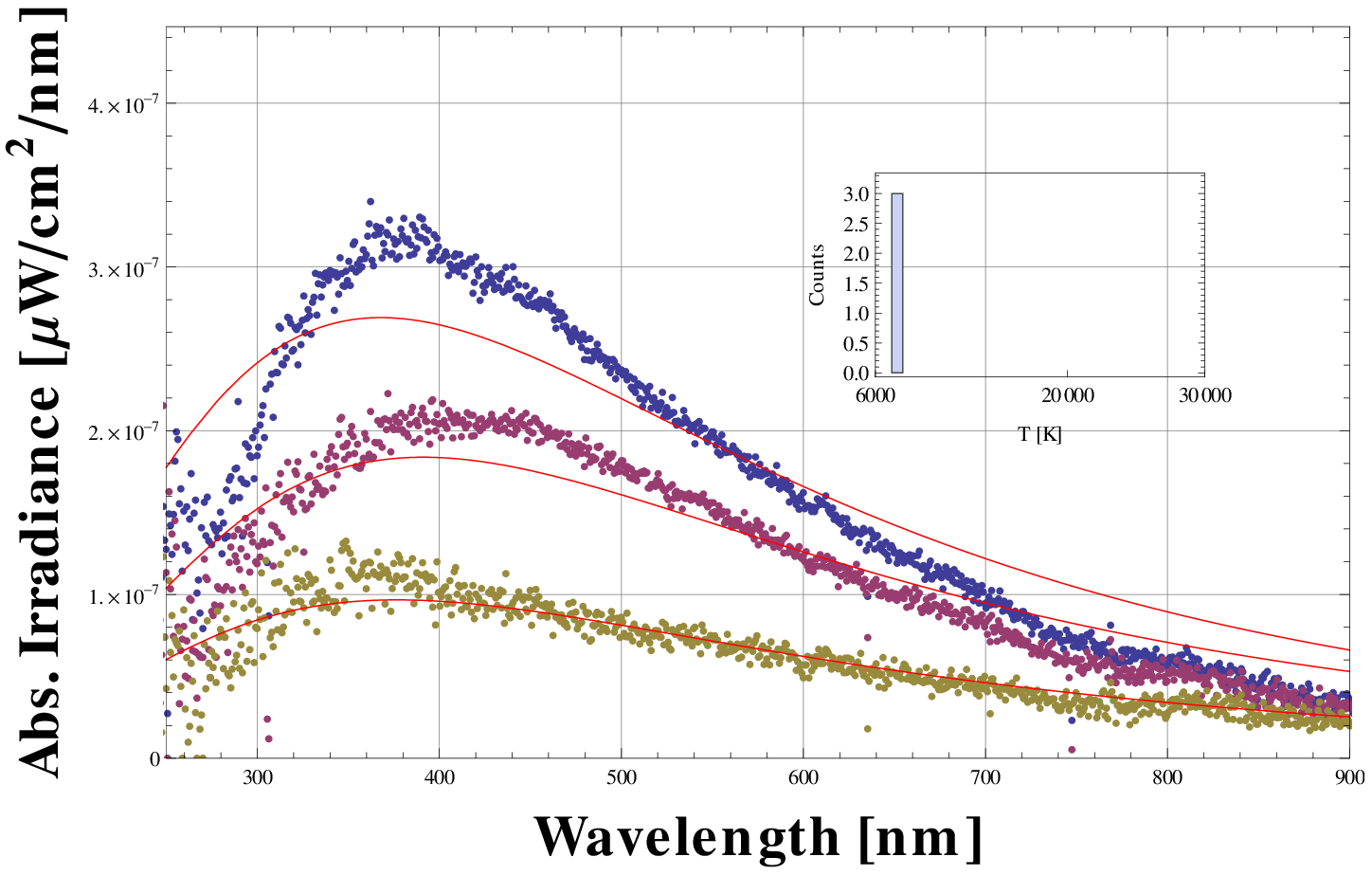} &
\includegraphics[width=2.5in]{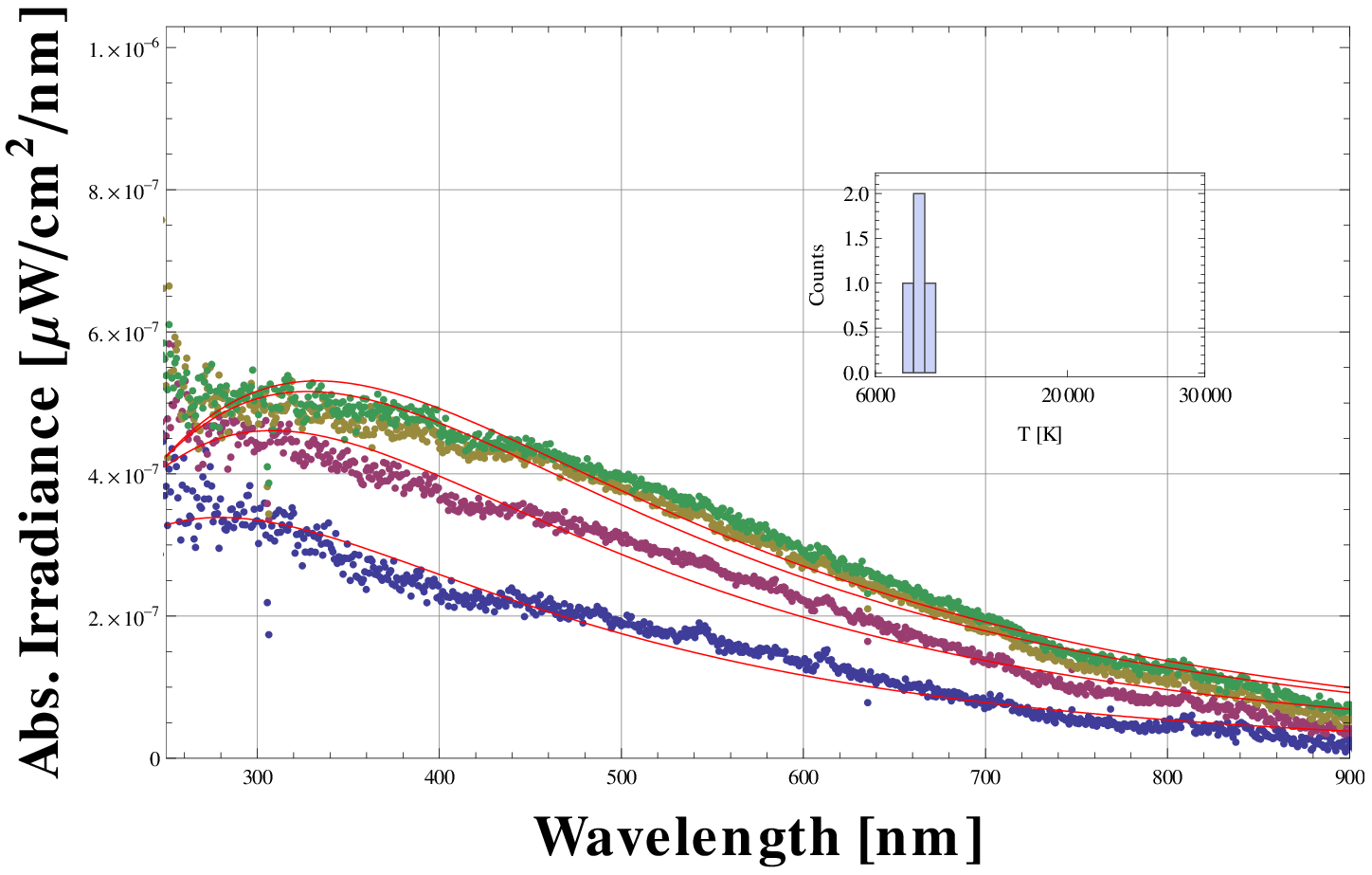} \\
\includegraphics[width=2.5in]{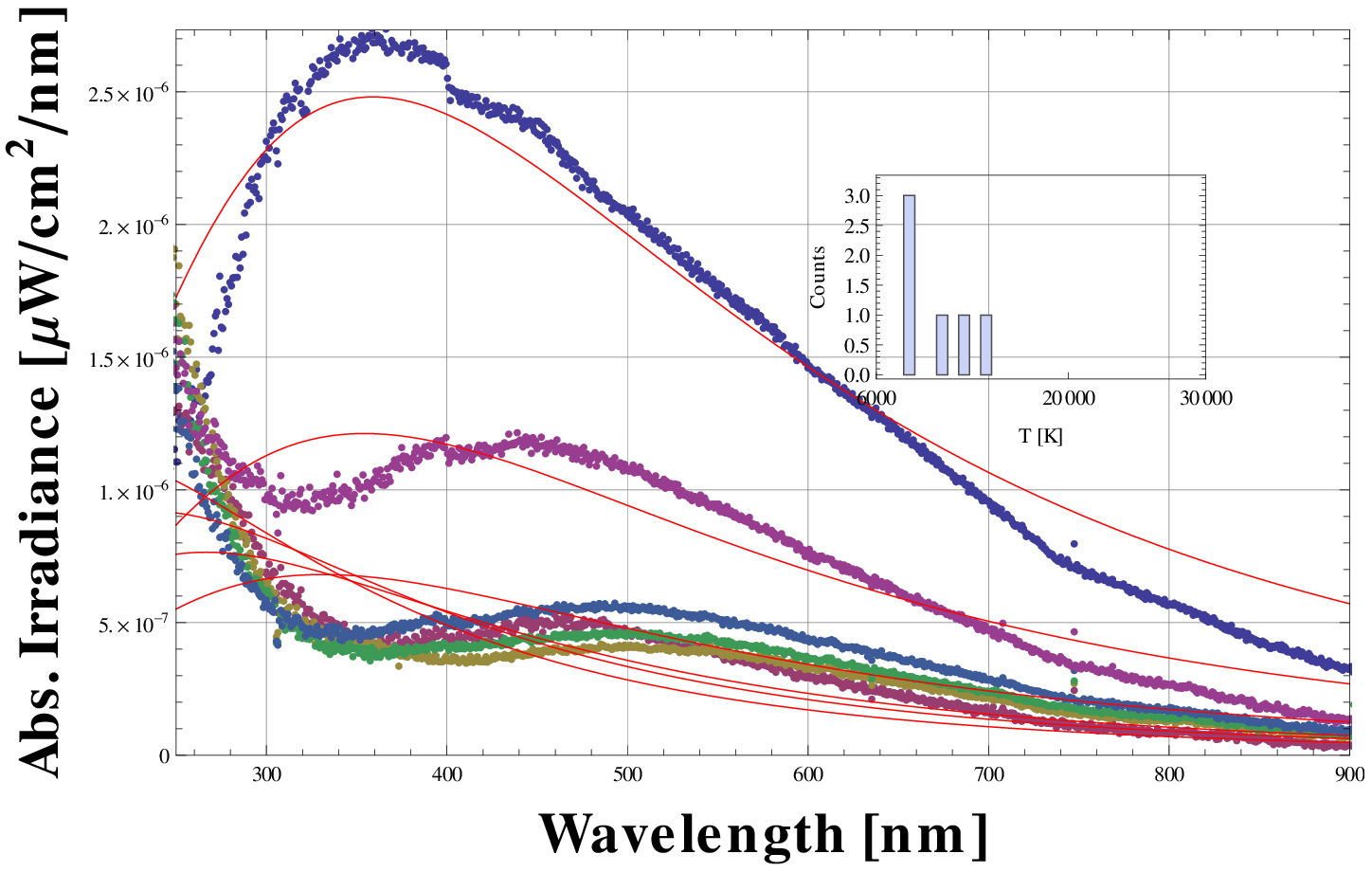} &
\includegraphics[width=2.5in]{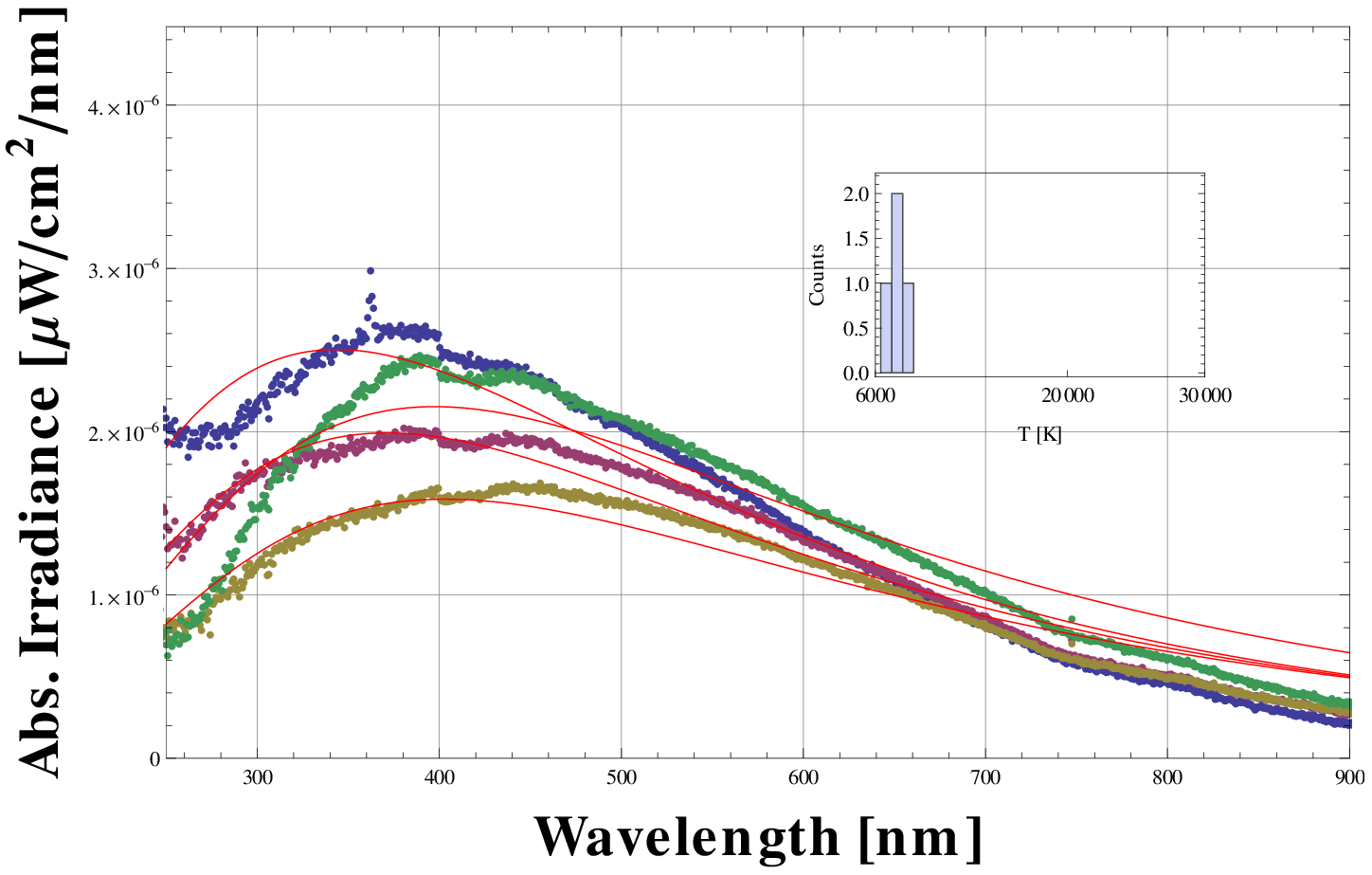} \\
\includegraphics[width=2.5in]{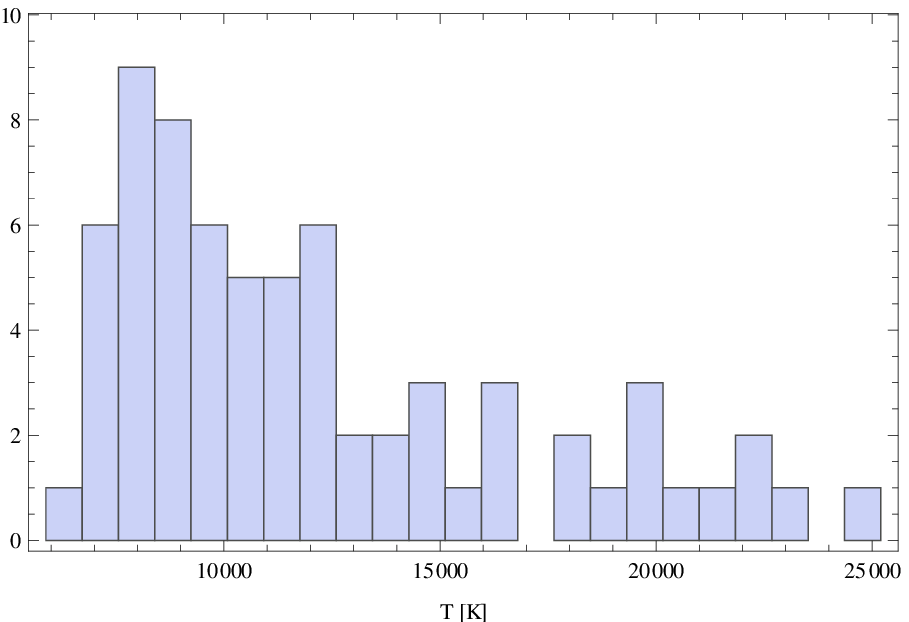} \\

\end{array}$
\end{center}
\caption{Another set of  measurements of spectra. Mean of temperature for above plots from top to bottom and from 
left to right is 7550, 9325, 10338, 7683 K. Plot at bottom represents fitted temperature distribution based on all spectra taken.  
Mean of fitted temperature is 12138 K.} 
\label{UVabs3}
\end{figure}

   Integration time for acquiring spectra was 60 s, TE cooling was applied \cite{ooqe65}. At these conditions QE65000 spectrometer has
   very stable response with
   dark current pretty much constant.\\
    In Figs. \ref{UVabs1}-\ref{UVabs3} there are collected all spectral measurements done for SBSL produced inside quartz resonator. 
 Spectra displayed are taken consecutively in time. Created sets in specific plots are chosen for purpose to emphasize either adherence
 to black-body radiation like shape or to emphasize differences in this direction. 
  One should mention that sub-sample of the sample presented  in Fig. \ref{UVabs1} was also used for a
  calibration of the spectrometer \cite{calibSL}. It was not a case for other plots.

\clearpage
\subsection{Conclusion}
 Spectra of SBSL  in UV - VIS range (250 - 900 nm) corrected for spectrometer bias have been studied. Generally, 
 spectra resemble black-body radiation shape for temperatures between 7000 - 20000 K. This is in accord with an old experimental and 
 theoretical knowledge
 (e.g. \cite{brenn}).\\
 In our measurements  we find notable differences from black-body radiation shape.  By analyzing these differences, as an explanation of
 quite unusual shapes of measured spectra,
 one tends to
 speculate about existence of two radiators - one at higher and one at lower
 temperature (see e.g. Fig. \ref{UVabs2}, plots in middle) which can be separated in time during burst of light. This would be in accord with time resolved measurement in \cite{timeres} and suggests 
 that time evolution of the spectra from SBSL as measured in \cite{timeres} are 
 not restricted to a specific case  but are general property of SBSL. At this stage we 
 also cannot  completely exclude 
 some ``apparatus'' effect. The topic needs more detailed study and better control of parameters SBSL depends on.
 
\end{document}